\documentclass{cernart}
\usepackage{epsfig}
\begin{document}

\def\He{^6{\mathrm{He}}}
\def\nutau{{\mathrm\nu}_{\mathrm\tau}}
\def\numu{{\mathrm\nu}_{\mathrm\mu}}
\def\nue{{\mathrm\nu}_{\mathrm e}}
\def\anue{\overline{{\mathrm\nu}}_{\mathrm e}}
\def\anumu{\overline{{\mathrm\nu}}_{\mathrm \mu}}
\def\dm2{\Delta\mathrm{m}^2}

\begin{titlepage}
\date{30-6-2001}
\title{A novel concept for a $\anue$ neutrino factory}
\begin{Authlist}
P. Zucchelli
\footnote{On leave of absence from INFN, Ferrara}
\Instfoot{a1}{CERN, Geneva, Switzerland}
\end{Authlist}
\begin{abstract}

The evolution of neutrino physics demands new schemes to produce intense,
collimated  and pure neutrino beams. The current neutrino factory concept 
implies the production, collection, and storage of muons 
to produce beams of muon and electron neutrinos at equal intensities at the
same time. 
 Research and development addressing its feasibility are ongoing.
In the current paper, a new neutrino factory concept is proposed, 
that could possibly achieve beams of similar intensity,
perfectly known energy spectrum and a single neutrino flavour,
electron anti-neutrino. The scheme relies on existing technology.

\end{abstract}                                            
\vspace{2cm}
\submitted{Submitted to Phys. Lett. B}
\Anotfoot{*}{Email address: Piero.Zucchelli@cern.ch}
\end{titlepage}

\section{Introduction}

The demand for better neutrino beams is correlated to the considerable 
improvement
in neutrino detectors, and to the recent exciting claims of evidence
 for neutrino oscillations by various experiments. 
In particular,
solar, atmospheric and accelerator neutrinos
 appear today to oscillate (and therefore
should have non-zero masses)  in a way that it is hard to accommodate 
in a unique picture, given current theoretical
understanding. Speculation and {\sl ad-hoc} theories abound in the absence
of decisive experiments. Obviously, 
a high intensity neutrino source of a single flavour, 
no background and perfectly known
energy spectrum and intensity 
could be decisive both for oscillation searches and precision
measurement of the lepton mixing parameters. 
 
\begin{figure}
\begin{center}
    \resizebox{0.8\textwidth}{!}{
      \includegraphics{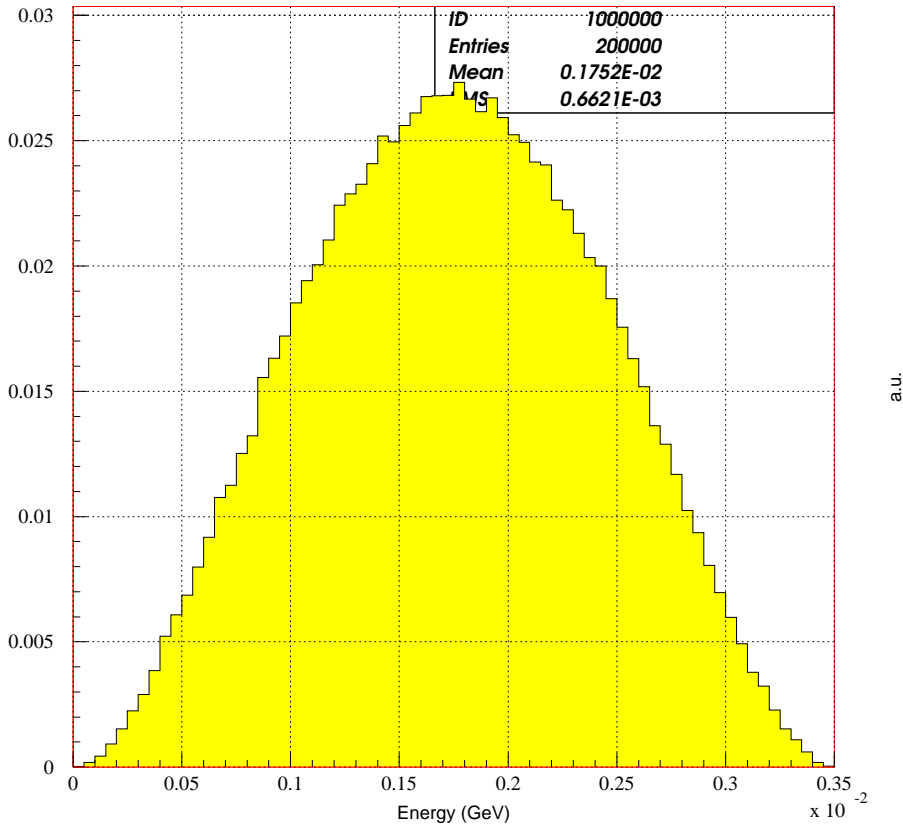}
        }
      \caption{Neutrino energy spectrum in the centre-of-mass frame for a $\He$ decay.}
        \label{betarest}
\end{center}
\end{figure}

\section{The concept}

It is proposed to produce a collimated $\anue$ beam by accelerating, at high energy,
radioactive ions
that will decay through a beta process. 

The radioactive ion production and acceleration to low energy (several MeV) 
have already been 
performed for nuclear studies, and various 
techniques have been developed \cite{isol},
e. g.  at CERN ISOLDE. 

The acceleration of the positively charged  atoms 
to about 150 GeV/nucleon is already done in the CERN PS/SPS 
accelerators for the heavy-ion programme.

The storage of the radioactive ion bunches in a storage ring could be very
similar to what is being studied for the `conventional' neutrino 
factory scheme \cite{garoby}. 

Two important features have to be outlined. 
\begin{itemize}

\item[$\bullet$] Unprecedented beams of single flavour neutrinos
with energy spectrum and intensity
known a priori. Moreover, this flavour ($\anue$) is different
from that which is dominant for conventional beams ($\numu$). 
This feature allows new precision measurements (oscillation searches 
 at small mixing angles, nuclear physics studies) to be performed.

\item[$\bullet$] The second peculiar characteristic is given by the fact that the neutrino
parent, the ion, has a  rest energy much larger than the neutrino energy
in the centre-of-mass frame.  This allows 
a focused beam of low-energy neutrinos to be produced, which has been
impossible up to now. This feature is particularly important
for long-baseline neutrino studies.
\end{itemize}
\begin{figure}
\begin{center}
    \resizebox{0.8\textwidth}{!}{
      \includegraphics{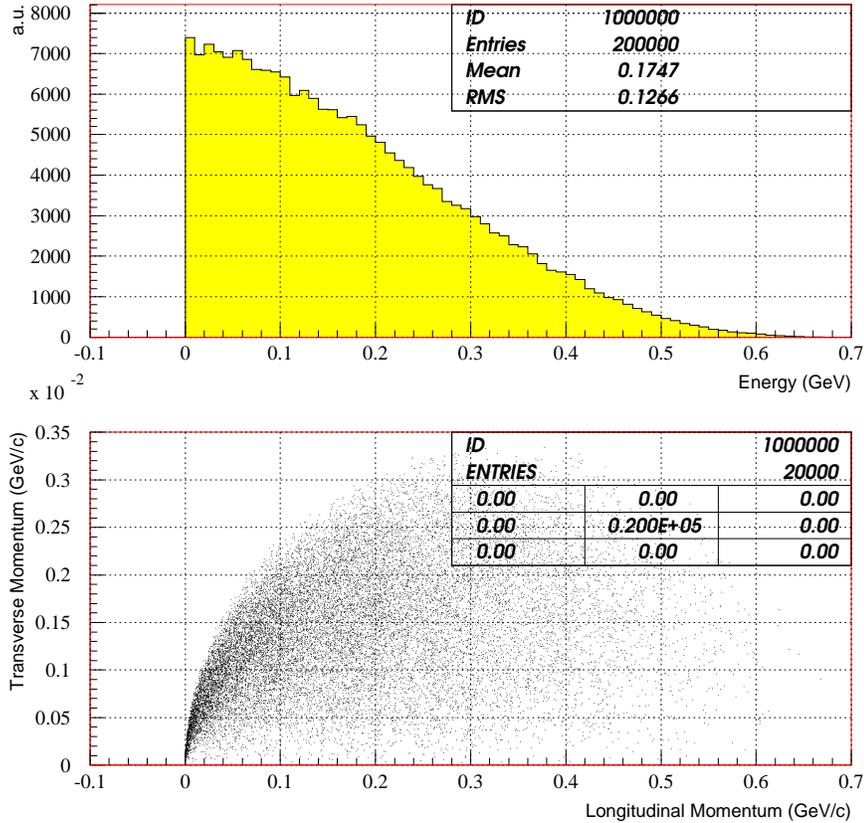}
        }
      \caption{`Boost' focusing of the neutrinos.} 
        \label{betaboost}
\end{center}
\end{figure}

\subsection{Nuclear beta decays}
As a guideline, a textbook atomic $\beta^-$ decay which has well-known 
characteristics and good features for neutrino production is considered:

$$ \He^{++} \rightarrow ^6_3\mathrm{Li}^{+++} \; \mathrm{e}^- \anue \;\;\;.$$ 

Its half-life $T/2$ is 0.8 s, and the beta decay endpoint $E_0$ (maximum energy of the emitted electron) 
is 3.5 MeV. The energetic endpoint of the electron 
and the atom lifetime are, unfortunately, correlated by the `Sargent rule' \cite{perkins}. In substance, the width of the unstable initial state is proportional to the fifth
power of the energy endpoint, so that a low $E_0$ value implies an almost stable atom. 
For neutrino production and long-baseline studies, contemporaneous low value of $E_0$ and $T/2$
would be the best solution, in contradiction with nature. A deeper
study of potential neutrino sources is mandatory but, for the sake of illustration, 
$\He$ is a valid starting point. 
The energy spectrum of the electron produced in the $\He$ beta decay has been extensively
measured and is well described theoretically (for energies larger than the electron mass and
valid without corrections only for light nuclei) by the simple analytic formula

$$ N(E) \mathrm{d}E \approx E^2(E-E_0)^2 \;\;\;,$$

where $E$ is the electron energy. 
The neutrino spectrum is completely known by the laboratory measurement of the associated
electron (without involving a neutrino measurement) 
since $E_\mathrm{e} +E_\nu \approx E_0$ because of the large 
mass of the nucleus. 
Fig. \ref{betarest} shows the neutrino
spectrum from $\He$ decays when the source is in the same 
reference frame as the experimenter.

\subsection{The relativistic effect}
Suppose that the $\He$ atom is accelerated up to a value of $\gamma = 100$, 
achieving a typical energy per nucleon currently obtained in the heavy-ion runs of the CERN SPS. 
In the laboratory frame, the neutrino transverse momentum (with respect 
to the beam axis) is identical to that observed when the atom is at rest: 
1.75 MeV on average. In contrast, the average longitudinal momentum is multiplied
by a factor corresponding to $\gamma_{\He}$ and therefore neutrinos have typical decay angles
of $1/\gamma_{\mathrm{PARENT}}$, in our case 10 mrad. Special relativity also tells us
that the neutrino energy in the forward direction is 
multiplied by twice the same factor, so that the average neutrino energy on 
a `far' detector is expected to be 350 MeV.

A better insight into the correlations among the various distributions can be obtained by 
a simple Monte Carlo simulation. The laboratory spectrum of all produced  neutrinos is shown in 
Fig. \ref{betaboost}, together
with the characteristic phase distribution in the $P_T-P_L$ plot that corresponds to the
relativistic transformation of an isotropic distribution in the rest frame.

\begin{figure}
\begin{center}
    \resizebox{0.8\textwidth}{!}{
      \includegraphics{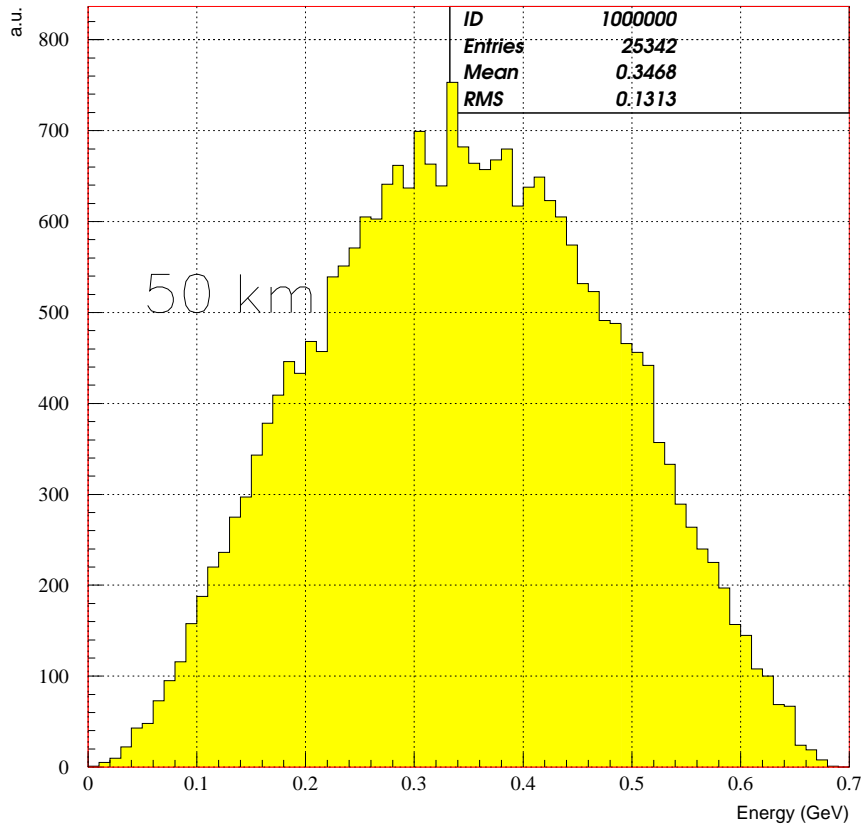}
        }
      \caption{Neutrino spectrum 50 km from the source.} 
        \label{betadist}
\end{center}
\end{figure}

The effective characteristics of this type of beam can be observed by looking in more detail 
at the neutrino
spectrum on an ideal detector in a far position. if the lateral dimensions of the detector
are smaller than $1/\gamma_{\mathrm{PARENT}}$ multiplied by the distance, the neutrino spectrum
has no radial dependence
and corresponds to the neutrino spectrum at rest, with an energy endpoint $E_0^{\mathrm{LAB}}=2\times \gamma \times E_0^{\mathrm{CMS}}$.
This has been verified by the simple Monte Carlo simulation and is shown, for one case, in Fig. \ref{betadist}.
The relative flux for various distances is reported in Table \ref{distance}.

As in the case of the spectrum, even the neutrino flux intensity can be 
evaluated from first principles:
the isotropy of the decay in the centre-of-mass
frame, the knowledge of the $E_0$ endpoint value and the acceleration energy ($\gamma$).

It is important to compare the focusing properties with those of a conventional neutrino factory
beam. The comparison should be done 
for identical values of $\left<E\right>/L$: a value of $ 7\times10^{-3}$ GeV/km is arbitrarily chosen, which corresponds 
to a 50 km baseline for the beta-beam and 4857 km for the conventional 
neutrino factory beam produced by 50 GeV muons \cite{garoby}. 
The flux of neutrinos reaching the far detector is $1.4\times 10^{-6}/ \mathrm{m}^2$ in the case of the beta-beam and 
$8.0\times10^{-9}/\mathrm{m}^2$ in the case of the conventional neutrino factory beam. After this, it has
to be said  that the collection efficiency comparison is essentially independent of the $\gamma$ factor in both
cases, if the comparison is made under identical $\left<E\right>/L$ conditions. 

\section{Feasibility}
Here, the challenges to the feasibility of beta-beams  are reviewed aspect
by aspect, and possible objections are anticipated.
\subsection{Radioactive ion production} 

Various techniques have been developed in the nuclear physics community 
in order to produce
unstable, radioactive nuclei. For a detailed review, the interested reader is
referred to ref. \cite{isol}. 
Probably, 
the technique developed at CERN ISOLDE, called ISOL ion production, is the most
suitable for a high  
intensity $\He$ production. Today ISOLDE can produce up to $\approx 10^{8}$ $\He$
ions per second \cite{wwwisolde}, without a specific optimization for this atomic state. A significant
improvement can be achieved in various ways: increased target thickness, optimized target converter, use of an ECR ionization source,  
higher proton energy \cite{nilsson}. 
It should be mentioned that a possible upgrade of the CERN 
accelerators by a Superconducting Proton Linac
(SPL) would additionally improve the production rate, and such a facility has already been advocated 
and studied by the ISOLDE
team \cite{isolteam} for nuclear physics experiments. 
The consequent further increase of proton energy and intensity could allow the 
production rate to be improved by two or three orders of magnitude,
achieving the value of $10^{12}/\mathrm{s}$ $\He$ ions which is 
an important ingredient for an effective beta-beam.

\subsection{Post-acceleration} 
Post-acceleration of $\He$ ions does not differ in the substance from non-radioactive ion acceleration.
Unlike the conventional neutrino factories, the acceleration time to reach the relativistic
regime (where the dilatation of the decay time in the laboratory frame occurs) is not critical, 
being the ratio of the lifetimes $4\times10^{5}$, and
for example the acceleration cycles of the CERN PS multi-purpose synchrotron are affordable. 
REX-ISOLDE is indeed a facility under commissioning for post-acceleration of a wide spectrum of radioactive
ions.
Once the relativistic regime is achieved (PS), the injection into a larger machine, CERN SPS for example, 
could take the beam to the required energy before it is sent to the storage ring.

Given that the beam is radioactive, one could object that the radioactive pollution in the accelerator
would seriously compromise the beta-beam's feasibility. In fact, 
the radioactivity rate could be tolerable because 
its characteristics are
significantly different from those of pure losses, and moreover it decreases at high energy because of the 
relativistic time dilatation due to the energy boost. 
Without giving a proof, obtainable only by a detailed
study on the specific accelerator,  a possible argument in favour of it
is presented below.

\begin{table}
\begin{center} 
\caption{Relative flux from a $\gamma=100$ ($\left<E_\nu \right>=350$ MeV) $\He$ beta-beam at various distances onto 
a detector of 1 $\mathrm{m}^2$. Corresponding
$\left<E\right>/L$ values are also reported.}
\label{distance}
 \begin{tabular}{|l|l|l|}
\hline
\hline
Distance (km)   & Relative flux ($\nu/\mathrm{m}^2/{\rm ions}$) & $\left<E\right>/L$ (GeV/km) \\ 
\hline
1     & $3.0\times10^{-3}$ & $0.35$ \\ 
3.125    & $3.5\times10^{-4}$ & $0.11$ \\ 
6.25    & $4.4\times10^{-5}$  & $5.6\times10^{-2}$ \\ 
12.5    & $1.7\times10^{-5}$ &  $2.8\times10^{-2}$ \\ 
25    & $5.2\times10^{-6}$   &  $1.4\times10^{-2}$ \\ 
50    & $1.4\times10^{-6}$    &  $7.0\times10^{-3}$ \\ 
100    & $3.8\times10^{-7}$ &    $3.5\times10^{-3}$ \\ 
200    & $8\times10^{-8}$ &    $1.7\times10^{-3}$ \\ 
400    & $2\times10^{-8}$ &    $8.7\times10^{-4}$ \\ 
 \hline
  \end{tabular}
\end{center}
\end{table} 

A beta-beam induced
radioactivity is a perfectly known process, where an accelerated ion nucleus decays into a
harmless neutrino, a negative electron, and an ion whose charge differs
by one unit from that of its parent. The decay rate
decreases linearly with the increasing energy because of the boost, therefore
the average decay occurring during acceleration has a low-energy parent and is in the low-energy section of the accelerator. 
The average energy of the escaping electron
is about 300 times lower than that of its parent, while the lithium ion has of course the same energy as the
decaying helium. The electrons can be easily
stopped by thin metallic shields, and produce no secondary neutrons. 
The lithium ions  have a different charge.
Their trajectories will therefore change and follow
different and well-known orbits; the collisions of the decayed nuclei will thus occur
in well-localized places.
Therefore it is probably possible  to shield the machine where the trajectories 
of the daughter ions are expected.

\subsection{The storage ring} 
The storage ring must:
\begin{itemize}
\item[$\bullet$] have a straight section whose length, relative to the total length, is as long as possible;
\item[$\bullet$] store the maximum number of bunches, to allow the ions time to decay;
\item[$\bullet$] be immune from the radioactive ion decays.
\end{itemize}

The first two of these requirements are similar to those of a conventional neutrino factory 
storage ring, 
and detailed studies have already been performed. It is assumed to be 
reasonable 
to store 140 
bunches
with a relative length of the straight section towards the detector of 28.7\% \cite{garoby}. 
The immunity to radioactive decays
can be obtained in the same way as discussed for the accelerating machines. As an interesting possibility,
 it is noted that at the end of the straight sections the intensity of lithium  ions
and electrons is maximum: A dedicated dipole-like optics could separate the electrons from the 
helium and from 
the lithium. Electrons could provide a very direct intensity monitor.  
The high-energy lithium ions could be recycled to activate the ion source and 
therefore improve the efficiency of the radioactive ion production. 

\subsection{Baseline, energy and intensity considerations} 

Order-of-magnitude performances of the acceleration scheme can be based on current efficiencies of 
existing machines. A loss rate of 50\% in the accelerator and a 140 s storage time in the storage ring are assumed, but the neutrino interaction
rate is {\bf not} included.

\begin{table*}[th]
\begin{center} 
\caption{Summary of possible performances and characteristics of a beta-beam.}
\label{beta-beam}
 \begin{tabular}{|l|l|l|}
\hline
\hline
$\He$ ions production  & $10^{12}$/s & \\
$\He$ accelerator efficiency& 50\% & \\
$\He$ final energy & 100 GeV/nucleon & \\
\hline
Storage ring bunches & 140 & \\
Straight section relative length & 28.7 \% & \\
Storage time & 140 s  & \\
\hline
Running time/year & $10^7$  s & \\
\hline
Neutrino flux at 1 km  & $3\times10^{15}/\mathrm{m}^2$/year & $\left<E\right>/L=0.3$ GeV/km (LSND) \\
Neutrino flux at 12.5 km & $1.7\times10^{13}/\mathrm{m}^2$/year & $\left<E\right>/L=2.8\times10^{-2}$ GeV/km (CNGS) \\
Neutrino flux at 25 km & $5.2\times10^{12}/\mathrm{m}^2$/year & $\left<E\right>/L=1.4\times10^{-2}$ GeV/km (NuFact) \\
Neutrino flux at 50 km  & $1.4\times10^{12}/\mathrm{m}^2$/year & $\left<E\right>/L=7.0\times10^{-3}$ GeV/km (Super-beam) \\
Neutrino flux at 100 km & $3.8\times10^{11}/\mathrm{m}^2$/year & $\left<E\right>/L=3.5\times10^{-3}$ GeV/km (SuperK Atm) \\
 \hline
  \end{tabular}
\end{center}
\end{table*}

A flux comparison with existing beams, proposed super-beams, and expected neutrino factory performances 
allows an estimation of whether the neutrino flux achieved by a beta-beam could have 
a significant impact on neutrino physics understanding. The reader is reminded that the CERN 
Neutrinos to Gran Sasso
(CNGS) beam aims at an integrated flux of $3.5\times10^{11} \; \numu/\mathrm{m}^2/$year
with a  17.7 GeV average energy \cite{elsener}\cite{ball}.
One of the discussed super-beam options \cite{mezzetto}
has the goal of $2.4\times10^{12} \; \numu/\mathrm{m}^2/$year  at 260 MeV. 
The conventional neutrino factory from muon decay aims to
$2.4\times10^{12} \nu/\mathrm{m}^2/$year at 34 GeV. 
Table \ref{beta-beam} shows the possible neutrino fluxes of a beta-beam at comparable 
$\left<E\right>/L$ values of physics relevance. 

The reader is explicitly warned against making comparisons that do not take into account the
physics objectives.
At first glance, for example, the super-beam and the beta-beam have similar
energy and similar intensity.  
But the different neutrino flavour of the beta-beam has a significant impact
on the main super-beam
physics objective \cite{cadenas}: 
the mixing angle $\theta_{13}$ can in fact be measured in the beta-beam by a 
direct disappearance
experiment or by neutral-current to charged-current analysis. Then, 
with a much lower background, a better knowledge of the neutrino flux and  a better knowledge
of the neutrino spectrum, the beta-beam can also perform in a better way 
the model-dependent
measurement based on the appearance of $\anumu$ from $\anue$ through indirect mixing with the third family, 
the beam related uncertainties being dominant \cite{cadenas} in the case of the super-beam.


\section{Impact on possible measurements}
To evaluate the physics impact of a beta-beam, the physics goal has to be specified in view of
the low maximum energy, the focusing property  and the different neutrino flavour with respect
to other accelerators.

\subsection{Oscillation physics: appearance and disappearance}

Disappearance measurements are particularly attractive since both intensity and spectrum
of the source are perfectly known on the basis of a non-neutrino measurement. 
These disappearance experiments 
have the advantage of being sensitive to oscillation
also if the value of $\Delta m^2$ is much larger than the typical $\left<E\right>/L$ (hereafter called normalized energy) of the experiment. 
For the case in which $\Delta m^2$ is comparable to the normalized energy, the experiment is ideally suited
to a precision measurement of the $\anue$ disappearance, with a sensitivity only limited by statistics.
When $\Delta m^2$ is smaller than the normalized energy, i.e. the experiment is too near to
the source,  the sensitivity of a disappearance experiment is typically seriously compromised and does not
appear to be competitive with appearance detection for the same normalized energy. 

A different and important evaluation of the oscillation parameters can be performed by studying
the ratio of neutral-current to charged-current interactions. In fact  the  study of 
$\nue \rightarrow \nutau$ oscillations is particularly difficult nowadays, since there are
only a few 
available $\nue$ or  $\nutau$ sources\footnote{The small $\nue$ and $\nutau$ contributions in conventional beams are not considered, 
since they are present only for high-energy neutrinos and therefore do not allow
a significant exploration of the region of $\Delta m^2$ below a few $\mathrm{eV}^2$, which is nowadays the focus
of the attention.}  
(solar and reactor neutrinos), which have furthermore a very low energy  and peculiar characteristics
(non-pulsed time structure, for example).

In conclusion, a disappearance beta-beam experiment could be a very large, simple electromagnetic 
calorimeter capable
of measuring the energy of one electron and located at a distance that makes the normalized 
energy comparable to the $\Delta m^2$ value to be measured. This detector, 
by timing, could be synchronized to the pulsed structure of the storage ring in order to minimize
backgrounds. It is impossible not to think of the large water Cherenkov detectors, such as 
SuperKamiokande or ANTARES.

Appearance experiments with beta-beams probably have to be limited to muon neutrino appearance. 
Even if it was possible to increase the ion  energy 
to achieve the cross-section threshold necessary
for tau production ($\nutau$ appearance), this would require substantial evolution of the 
accelerating technology: Machines with the characteristics
of LHC, for example,  employ sophisticated superconducting magnets which -- probably -- are not suited
to the larger ionization loss in the optical elements of the ring. The boost effect,
in addition, would imply a large storage capacity of the accumulator due to the additional lifetime dilatation.

Anyway, $\numu$ appearance experiments with beta-beams have two interesting possibilities
connected to the absence of other flavours in the beam. 

The first is related to the long-baseline measurements, since the far detector  
can be much simpler than a conventional neutrino factory detector, 
and it should only differentiate a muon track from an electron
shower. This simplicity could be reflected in a larger overall mass than that 
of a magnetic
detector  typically required in a conventional neutrino factory scheme to separate
the huge background induced by neutrinos of the same flavour but opposite lepton number.
Therefore, the ideal beta-beam appearance detector is a very large, simple 
detector with good muon identification properties. 
Again, it is impossible not to think of the large water Cherenkov detectors, 
as already studied in ref. \cite{cadenas} in the same energy regime.

The second possibility is connected to short-baseline measurements: 
If the MiniBoone experiment validates the LSND oscillation claim, 
a beta-beam experiment looking to $\anue \rightarrow \anumu$ 
oscillation without backgrounds could allow unprecedented measurements of 
oscillations in the region of $\Delta m^2$ relevant to astrophysics
and cosmology.  At the moment, no pure sources of $\numu$ or $\nue$ are available to 
appearance experiments which have to explore the domain of $\sin^2\theta_{12}\approx 10^{-4}$.

\subsection{Precision physics}

The high intensity and the absence of backgrounds make a beta-beam very interesting in the domain
of nuclear studies with neutrinos. The small maximum energy, however, 
limits the scope of the investigations to this domain, which could be very relevant anyway 
in order
to measure the cross-section on different target materials of $\anue$ neutrinos from astronomical sources.
 
\subsection{Are CP violation measurements still possible?}
 
An evident advantage of the conventional neutrino factory beams is the possibility to
 accelerate
muons or anti-muons to produce neutrinos of opposite lepton number. This opens the possibility to measure
CP violation effects in the  leptonic sector. 
Despite the difficulty in identifying the experiment parameters which will allow an effect
to be observed, 
this measurement is clearly of maximum importance. 
A beta-beam has 
the crucial advantage of a lower energy and better focusing, which is reflected in a larger
explorable domain of normalized energy values (see Table \ref{distance}).
But anti-helium atoms are impossible to produce. 
The solution is to accelerate a different atom, which 
has a superallowed $\beta^+$ transition \cite{wu}; for example,

$$ ^{18}_{10}\mathrm{Ne} \rightarrow ^{18}_{9}\mathrm{F}^{-} \; \mathrm{e}^+ \nue \;\;\;, $$

which has a half-life $T/2$ of 1.6 s and an endpoint $E_0=3.2$ MeV. 
Other candidates exist; for example, $^{34}_{17}\mathrm{Cl}$ and $^{38}_{19}\mathrm{K}$
have short lives and produce $\nue$ in their beta decay.

\section{Conclusions}

An alternative neutrino factory scheme can produce $\anue$ beams from the beta decay 
of boosted ions. The efficient focusing makes it very suitable for explorations at low $\Delta m^2$
values. The unprecedented beam flavour, the known spectrum, and the perfect purity make it attractive 
for both appearance and disappearance oscillation experiments, and for new precision neutrino
physics. The technology to produce, accelerate and store radioactive 
ions has already been explored.

\par

Neutrino physics recent discoveries probably represent the most important opportunity
to probe the Standard Model understanding. Any improvement of the current artificial
neutrino acceleration technology should therefore be investigated. Up to now, 
many have considered 
focused low-energy neutrino beams impossible. 
\par

\section{Acknowledgements}
The author is indebted to T. Nilsson, J. Panman and B. Saitta for their attention, 
comments and suggestions.

\end{document}